\documentclass[aps,prl,amsmath, twocolumn, groupedaddress]{revtex4}
\usepackage{amstext}
\usepackage{graphicx}
\usepackage{amssymb}

\makeatletter

\providecommand{\LyX}{L\kern-.1667em\lower.25em\hbox{Y}\kern-.125emX\@}
\providecommand{\tabularnewline}{\\}

\makeatother

\begin{document}

\preprint{}

\title{Fast adiabatic qubit gates using only $\sigma_z$ control}

\author{John M. Martinis}

\affiliation{Department of Physics, University of California, Santa Barbara, CA 93106, USA}

\author{Michael R. Geller}

\affiliation{Department of Physics and Astronomy, University of Georgia, Athens, Ga 30602, USA}

\date{\today}

\begin{abstract}
A controlled-phase gate was demonstrated in superconducting Xmon transmon qubits with fidelity reaching 99.4\%, relying on the adiabatic interaction between the $|11\rangle$ and $|02\rangle$ states.   Here we explain the theoretical concepts behind this protocol that achieves fast gate times with only $\sigma_z$ control of the Hamiltonian, based on a theory of non-linear mapping of state errors to a power spectral density and use of optimal window functions.  With a solution given in the Fourier basis, optimization is shown to be straightforward for practical cases of an arbitrary state change and finite bandwidth of control signals.  We find that errors below $10^{-4}$ are readily achievable for realistic control waveforms.
\end{abstract}

\maketitle

\section{introduction}

High fidelity controlled-phase (CZ) gates were recently demonstrated
for adjacent qubits in a 5-qubit quantum processor \cite{Rami13}. For
a two-qubit gate, CZ fidelity as high as 99.4\% was obtained using
the phase shift from near crossing of the $|11\rangle$ and
$|02\rangle$ states, similar to previous work on superconducting
qubits \cite{cz1,czad,cz2,cz3}, but following a new ``fast
adiabatic'' protocol that gives both fast gate times and low error
rates.  Here we explain the theoretical concepts behind this
protocol, showing how waveforms may be generated and parameterized in
a manner that is readily optimized, both numerically and
experimentally.

The general problem of achieving fast adiabatic performance is of great interest to the physics community
\cite{gen1,gen2,gen3, gen4}, with applications in coherent manipulation, precision measurements, and quantum computing.  Finding an improved control methodology therefore has potential for applications in a variety of quantum systems.

Although adiabatic quantum control has a long history, we are not aware of a theory that could address in detail our particular experimental problem: what is the optimal control trajectory for the desired CZ operation, having both short gate time and small non-adiabatic error, described in a way that is readily optimized with numerical and experimental methods.  For example, although a previous experiment used adiabaticity for a CZ gate \cite{czad}, the characterization and optimization of the adiabatic control waveform was not developed.  The goal here is to describe a simple theory that can be used to intuitively understand optimal waveforms for fast and accurate operation.

Since we are concerned here with transition errors between the states $|11\rangle$ and $|02\rangle$ as their frequencies are varied, we may map this to a qubit problem for states $|0\rangle$ and $|1\rangle$ that has a fixed energy in $\sigma_x$ from the avoided-level crossing and adjustable qubit frequency in $\sigma_z$.  We note that this optimization problem differs from conventional quantum control, where the $\sigma_z$ part of the Hamiltonian is fixed and both the $\sigma_x$ and $\sigma_y$ parts can be varied.  Here, the single-component control in $\sigma_z$ makes optimal control much more difficult since a second operator does not exist that could be used to restore the desired adiabatic character \cite{demirplak}, disallowing standard pulse shaping techniques \cite{demirplak, motzoi} and ``superadiabatic'' theory \cite{super0,super1,super2}.

The set of possible adiabatic solutions is also influenced by other constraints with the superconducting qubit system.  First, the non-linearity of the qubits is small \cite{Rami13}, so other unwanted transitions are not necessarily far off resonance.  For example, although the waveform is optimized for adiabaticity of the near resonance $|11\rangle$ and $|02\rangle$ states, fast changes in the control waveform may produce transitions between the states $|01\rangle$ and $|10\rangle$, even though off resonance by $\gtrsim 200\,\textrm{MHz}$.  A natural solution to this constraint is using smooth waveforms, which also seems to be a good strategy in terms of controlling many qubits, achieving low crosstalk, and simplifying the requirements of control electronics and their system calibration.  We thus rule out pulse and refocussing type protocols common in nuclear magnetic resonance (NMR) \cite{nmrpaper,steffen}, but welcome any theory making this a practical solution.

Second, we are looking for control waveforms with fast gate times, of order $40\,\textrm{ns}$ for our qubits, that efficiently uses the moderate coupling $g/2\pi \simeq 30\,\textrm{MHz}$ between qubits \cite{gdef}; note that $2\pi/g = 33\,\textrm{ns}$, making this problem non-trivial for adiabaticity.  Fast gates are important for coherence, especially since errors from 1/f flux noise and off-resonant coupling to other qubits, described as frequency shifts, produce phase errors proportional to characteristic gate times: If limited by these processes, the error probability thus scales as the \textit{square} of gate time.  A full system optimization of a qubit processor, for example describing the optimal choice of on-resonance qubit-qubit coupling $g$ and its smaller off-resonant coupling, is beyond the scope of this article.  But clearly even a small improvement of gate time can be quite important, motivating here a detailed understanding of both fast and accurate control.

Third, the CZ gate uses adiabatic control that is non-standard, as it partially moves into the avoided level crossing regime and then back out again, with the amplitude and time of the waveform adjusted to give a $\pi$ phase shift to the $|11\rangle$ state.  Typical solutions to an adiabatic problem consider moving through an avoided-level crossings \cite{szpaper,qs,ad1}, so a theory is needed to calculate optimal waveforms for an arbitrary control problem.

Since one varies only a single control parameter in $\sigma_z$, this theory may be described in a simple manner with familiar concepts that give insight into a physical understanding of adiabatic behavior.  We first show how state change from $\sigma_z$ control in the qubit Hamiltonian can be intuitively described in an instantaneous coordinate system.  Here, the evolution of the quantum system can be understood in the small error limit using a geometrical construction.  Second, this description allows errors to be calculated using a generalized Fourier transform.  Third, to obtain a minimum time solution for small changes in angle, the theory of window functions is used to find optimal control waveforms.  We then show how to efficiently parameterize the optimum waveform using a few-term Fourier series.  Lastly, a non-linear mapping is used to find the optimal control waveform for an arbitrary change in the control.

We also examine a second problem, microwave control of a qubit, and show that conventional two-component optimal control using $\sigma_x$ and $\sigma_y$ gives better results as expected, but not by much.

These results present a simple yet effective way to optimize for adiabatic fidelity when changing control only in $\sigma_z$.  Perhaps surprisingly, non-adiabatic errors can be small even if the interaction time is only slightly larger than the time scale of the transverse coupling, similar to that proposed earlier for a non-adiabatic gate \cite{gellercz}.  A recent experiment by our group has also demonstrated that these theoretical ideas can be used to optimize CZ performance \cite{kelly}.

\section{Geometrical Solution}

\begin{figure}[t]
\includegraphics[width=0.48\textwidth]{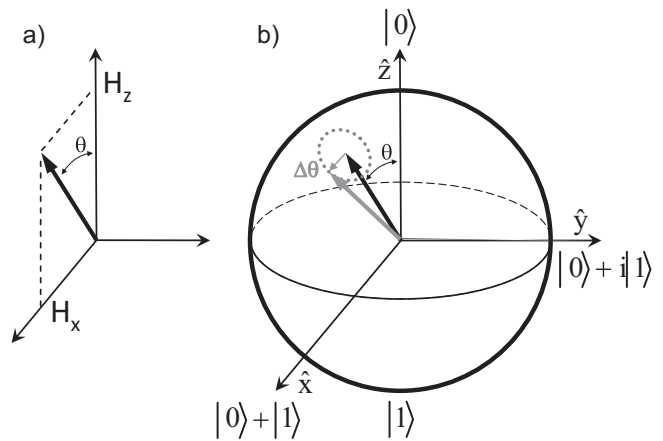}
\caption{\label{fig:bloch} Geometric picture of non-adiabatic errors.  a) Control field of constant $H_x$ and variable $H_z(t)$ gives control angle $\theta$ from z-axis.  b) Bloch sphere representation gives ground eigenstate parallel to control vector (solid arrow).  For a small step in $\theta$, the quantum state (gray arrow) points slightly away, with change $\Delta \theta$ (small gray arrow).  This state rotates around the eigenstate at frequency $\omega$ (gray dotted lines).  }
\end{figure}

We start with a geometrical solution to non-adiabatic error, a problem previously described using conventional methods \cite{super2,deschamps}.  The description of a qubit given by the 2-state Hamiltonian is
\begin{align}
H = H_x\sigma_x + H_z\sigma_z
  = \left(
      \begin{array}{cc}
        H_z & \ \ H_x \\
        H_x & -H_z \\
      \end{array}
    \right) \ ,
\end{align}
where $H_x$ is constant and the qubit frequency is controlled by a time dependent $H_z(t)$, for example the magnetic field in the z-direction for an electron spin.  We consider the qubit to be controlled by a Hamiltonian vector $(H_x,0,H_z)$, having both length and direction.  As shown in Fig.\,\ref{fig:bloch}a, we define the control parameter $\theta$ as the angle between the Hamiltonian vector and the z-axis
\begin{align}
\theta = \arctan(H_x/H_z) \ .
\label{eq:theta}
\end{align}

The qubit state is described by a Bloch vector that can point in any direction on the Bloch sphere.  Here, the usefulness of this geometric construction is the idea that the lowest energy eigenstate has a Bloch vector that points in the direction of the control vector: the eigenstate vector has angle $\theta$ from vertical and lies in the plane of the $\hat{x}$ and $\hat{z}$ axes.  The ground state eigenvalue $E_-=-\sqrt{H_x^2+H_z^2}$ is proportional to the length of the control vector, whereas the excited eigenstate points in the opposite direction with eigenenergy $E_+=-E_-$.  If the Bloch state vector is not an eigenstate, then its time dynamics is to rotate (precess) around the eigenstate vector at a frequency proportional to the difference in eigenstate energies
\begin{align}
\omega = (E_+-E_-)/\hbar =2\sqrt{H_x^2+H_z^2}/\hbar \ .
\label{eq:omega}
\end{align}

Non-adiabatic errors are typically understood by considering the Landau-Zener trajectory $H_z=\dot{H}_z t$.  Initially at large negative times $t$, where $H_z$ is large and negative, the Hamiltonian control vector points in the down direction, with the state given by the Bloch vector also in the down direction having $\theta = \pi$.  As the time goes to zero, the control vector  $(H_x,0,H_z=0)$ points along the equator, with the Bloch vector $\theta = \pi/2$.  At large positive times the control and Bloch vector points in the up direction $\theta = 0$.

When the field is slowly varied $(\dot{H}_z \rightarrow 0)$, the state adiabatically changes from $\theta = \pi$ to $\theta = 0$.  A finite ramp rate gives errors, with the probability to make a transition to the excited final eigenstate as
\begin{align}
P_e=\exp(-\pi H_x^2/\hbar\dot{H}_z) \ .
\end{align}
A small final error $P_e = 10^{-4}$ implies the condition $\dot{H}_z=0.341\,H_x^2/\hbar$.  If we assume a change in $H_z$ from $-10 H_x$ to $10 H_x$, the total time for the control pulse is $t_c = 18.6\,(h/2H_x)$, where $h/2H_x$ is the oscillation time for the transverse Hamiltonian.

The factor 18.6 implies a long control time for the Landau-Zener trajectory.  We will show with optimal control this time factor can be reduced significantly, to order unity.

For adiabatic evolution of the quantum state, the state vector stays aligned with the direction of the control Hamiltonian.  To better characterize non-adiabatic errors, we change the reference frame of the Bloch sphere to coincide with that of the control Hamiltonian.  In this moving frame $\theta_m=0$ represents being in the ground state of the instantaneous Hamiltonian, with no non-adiabatic error.

To understand non-adiabatic deviations, we consider the time dynamics of state change.  We first consider an infinitesimal time step $\Delta t$, during which the the control vector initially changes its direction by $\Delta \theta$, as shown in Fig.\,\ref{fig:bloch}b.  A transformation to the new moving frame produces a small change in the angle of the Bloch vector $-\Delta \theta$.  During the remainder of the time $\Delta t$ the Bloch vector rotates around the $\theta_m = 0$ axis, at the frequency $\omega$.  Starting with $\theta_m =  0$ before this time step, the angles $\theta_x$ and $\theta_y$ in the x- and y-direction can be written  compactly as
\begin{align}
\theta_m \equiv \theta_x+i\theta_y = -\Delta\theta \, e^{-i\omega \Delta t}
\end{align}
after time $\Delta t$, assuming small angles as appropriate for small non-adiabatic errors.  This finite angle implies the quantum state has not changed adiabatically (as $\theta_m \neq 0$), with the error probability given by $P_e = (\sin|\theta_m|/2)^2 \approx |\theta_m|^2/4$.

We next understand the error coming from a sequence of time steps $\Delta t_i$ each with a small angle change $\Delta \theta_i$.  A useful simplification comes from changing to a new frame that rotates with the precession of the Bloch vector, which implies the above $\theta_m$ vector does not rotate.  Instead, the change $\Delta\theta_i$ rotates in time by an angle $\phi_i$, which is computed by summing over all the previous phase changes in the individual steps. The net change in the complex angle of the Bloch vector in this moving and rotating frame is
\begin{align}
\theta_{mr} &= -\sum_i \Delta\theta_i \, e^{-i\phi_i} \\
&= -\int dt\,(d\theta/dt) \, \exp[-i \smallint^t \omega(t') dt'] \ , \label{GLint}\\
P_e&= |\theta_{mr}|^2/4 \ ,
\end{align}
where $P_e$ is the total probability error.  These equations are the general solution for non-adiabatic error in the small error limit, derived here using a geometric description.

The amplitude error $\theta_{mr}$ is proportional to the rate of change in the eigenstates, as given by $d\theta/dt$.  This quantity is mostly averaged to zero because of the rapidly changing phase from the $\omega$ integral, with averaged amplitude decreasing with increasing qubit frequency $\omega$.

Note that we have assumed here that all changes add linearly.  This is a good assumption as long as the net change in angle is very small; for example, it is known that for qubits with small change in state can be approximated by a harmonic oscillator, a linear system.  This assumption is generally acceptable since we are only interested in qubit control that give small errors.

An exact calculation of the error, starting from the Schr{\"o}dinger equation, is derived in Appendix I.  It gives results identical to this geometrical solution in the small error limit.  There are two changes for the exact solution: the driving amplitude $d\theta/dt$ should also be multiplied by the factor $\cos|\theta_{mr}|$, and the probability is $P_e=\{\sin[\arcsin(\theta_{mr})/2]\}^2$.

To understand how to construct an optimal waveform, we first note that there is a trade-off between the control time and the magnitude of non-adiabatic error.  Zero error is not possible - we are instead looking for acceptably small errors, say below $10^{-4}$, with as short of a control time as possible.  We also want stable control waveforms, so that the error does not increase rapidly for small changes in the waveform.

We first note that non-adiabatic errors are proportional to a change in the control variable $\theta$.  Errors will be relatively small when $|H_z| \gg H_x$, due to the slow dependence of $\theta$ on $H_z/H_x$ and the large oscillation frequency $\omega$, as described in Eqs.\,(\ref{eq:theta}) and (\ref{eq:omega}).

\section{Optimal Solution: small change in $\theta$}

For the case where $H_z$ only changes by a small amount, the qubit frequency can be approximated as being constant $\omega = \omega_0$.  Then Eq.\,(\ref{GLint}) is just the Fourier transform of $d\theta/dt$ at frequency $\omega_0$, which gives a probability proportional to the power spectral density of the signal at the oscillation frequency $\omega_0$
\begin{align}
P_e = (1/4)\, S_{d\theta/dt}(\omega_0) \ .
\end{align}
This makes sense physically since it is power at the transition frequency, given by $|\theta_{mr}|^2$, that drives the qubit transition to produce errors.

\begin{figure}[t]
\includegraphics[width=0.48\textwidth]{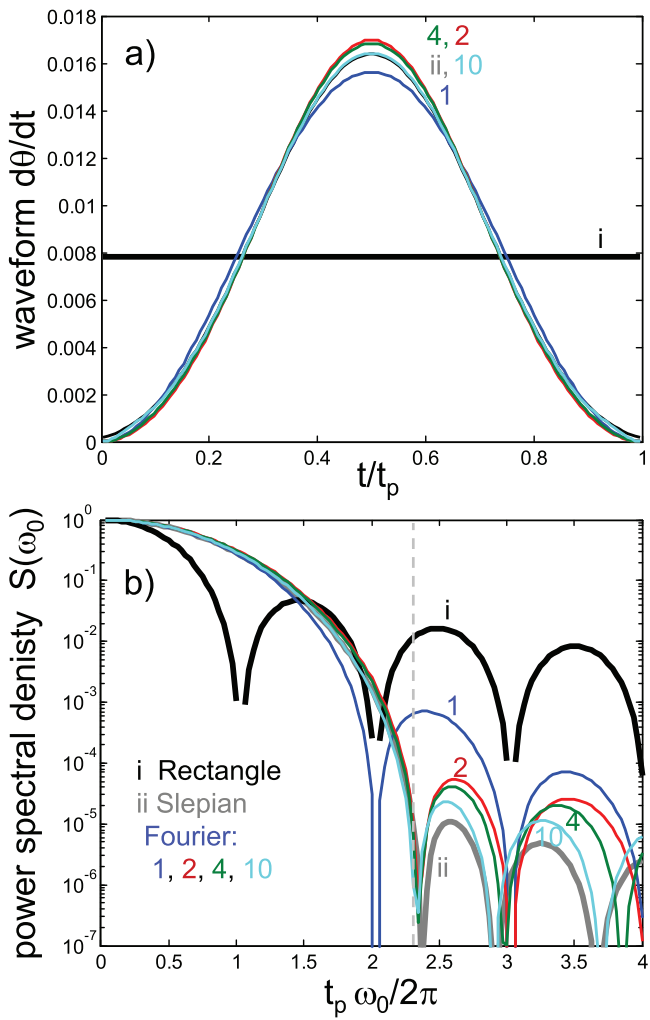}
\caption{\label{fig:Theta} (color online) Waveform and power spectral density $S(\omega_0)$, for a small change in control $\theta$ such that the qubit frequency is approximately constant $\omega = \omega_0$.  Here, the spectral density is dimensionless and proportional to the qubit error $P_e \simeq (1/4)S(\omega_0)$, valid in the limit of large $t_p$.  a) Plot of waveform $d\theta/dt$ versus $t/t_p$ for a rectangular pulse (i, black), Hanning window (1, blue), Slepian (ii, gray), and optimized Fourier coefficients with two (2, red), four (4, green), and ten (10, cyan) terms.  Note the similarity of the Slepian curves to the optimized Fourier functions, except near $t/t_p=0,1$.  b) Plot of power spectral density obtained from a Fourier transform of the waveforms in a). All waveforms are normalized to have unit area.  The vertical dashed line shows spectral cutoff for minimization of integrated noise.  }
\end{figure}

For the simple case of a linear change in the control $\theta$ from $\theta_i$ to $\theta_f$, the error is proportional to the (constant) derivative $d\theta /dt=(\theta_f - \theta_i)/t_p$, shown as the black line in Fig.\,{\ref{fig:Theta}}a.  The error versus the pulse time $t_p$ is the power spectral density of a rectangular pulse
\begin{align}
P_e^\textrm{sq} = (1/4)\Big|(d\theta/dt) \int_0^{t_p} dt \exp(-i\omega_0 t) \Big|^2 \\
= (\theta_f-\theta_i)^2 \sin^2(\omega_0t_p/2)/\omega_0^2 t_p^2 \ ,
\end{align}
which corresponds to the square of the sinc function.  Although the error is zero when $t_p = 2\pi/\omega_0$, small errors occur only around a small range of this time, so we consider this solution not to be stable or practically useful.  For large times, the general falloff of the error is quite slow, scaling as $1/t_p^2$, so this control waveform does not present a good solution to adiabatic control.

The functional dependence of the error versus pulse time can be changed by using different control waveforms \cite{warren}.  Another example is the function $[1-\cos(2\pi t/t_p)]/2$, known as the Hanning window, as shown in blue in Fig.\,{\ref{fig:Theta}}.  In this case, the error is
\begin{align}
P_e^\textrm{H}
= P_e^\textrm{sq} /|1-(\omega_0 t_p/2\pi)^2|^2\ ,
\end{align}
where the first zero is at twice the time of the square window $t_p=4\pi/\omega_0$, but the error falls off more quickly as $1/t_p^6$ at large times.  Although this window function is superior to the square window, the errors are still somewhat large right after the first zero, so other window functions should be considered.

As well known from NMR \cite{freeman}, optimizing non-adiabatic error versus total control time $t_p$ can be understood as optimizing windowing functions: the optimization criteria depend on the requirements for a particular application.

Since we are interested in low errors for a range of pulse times, we define the optimal waveform as one that gives low error for any time larger than some chosen time: this definition is essentially equivalent to adiabatic behavior.
We can use the theory of optimal window functions in signal processing to find this control waveform, known as a Slepian \cite{numrec}.  This function is defined by finding the optimal waveform that minimizes the integrated spectral density above a chosen frequency.  For example, the Slepian waveform optimized for frequencies above $\omega_0 t_p/2\pi > 2.3$ is shown in gray in Fig.\,{\ref{fig:Theta}}.  Although the first minimum is at slightly longer times than for the Hanning window, the error is always below $10^{-5}$ above this value.  A different Slepian cutoff parameter can be used to tradeoff the maximum error for the time of the first minimum.

A disadvantage of the Slepian function is that it does not go to zero at the beginning and end of the pulse, which creates a sharp edge as for a rectangular pulse.  This makes the function harder to synthesize, since real waveform generators have finite response times.  A practical way around this issue is to find a near optimum solution using Fourier basis functions \cite{freeman} given by
\begin{align}
\frac{d\theta}{dt} &= \sum_{n=1,2,\ldots, n_m} \lambda_n [1-\cos(2\pi n t/t_p)] \ , \label{eq:dth}
\end{align}
which goes smoothly to zero at the beginning and end of the pulse. The coefficients are constrained by the height of the pulse
\begin{align}
\theta_f-\theta_i &= t_p \sum_n \lambda_n \ .
\end{align}
Using standard numerical minimization (\textsf{fminsearch} function in MATLAB), the coefficients $\{ \lambda_n \}$ can be found that optimizes for the minimum integrated spectral density, as for the Slepian. The waveform and spectral densities for $n_m = 2, 4,$ and $10$ are plotted in Fig.\,{\ref{fig:Theta}}, which shows performance close to the Slepian.   Coefficients are given in Table\,\ref{tab:lamfourier}, showing the first two are dominant.  It is particularly interesting to note that the waveform for $n_m=2$ (red), with only 2 Fourier coefficients, is reasonably close in shape to the Slepian and has acceptably small errors $<10^{-4}$.

\begin{table}[t]
\caption{\label{tab:lamfourier} Table of $\lambda_n$ coefficients obtained from numerical minimization of noise for frequencies $\omega_0 t_p/2\pi > 2.3$, using the constraint $\sum \lambda_n =1$.}
\begin{tabular}{c|ccccccc}
\hline
\hline
$n_m$ & $\lambda_1$ &  $\lambda_2$ &  $\lambda_3$ &  $\lambda_4$
&  $\lambda_5$ &  $\lambda_6$ &  $\lambda_7$ \tabularnewline
\hline
2 & 1.0866 & -0.0866 & & \tabularnewline
4 & 1.0751 & -0.0811 & 0.0017 & 0.0044 \tabularnewline
10 & 1.0280 & -0.0606 &  0.0052 & 0.0055 & 0.0047 & 0.0046 & 0.0035 \tabularnewline
\hline
\hline
\end{tabular}
\end{table}

The Fourier basis functions can also include the terms $\lambda_n^s \sin(\pi n t/t_p)$ for $n$ odd, which go to zero at $t=0, t_p$ but have non-zero derivatives there.  Upon using this larger basis set, we find the numerically optimized solutions are not much better than found above only using cosine terms.  We choose not use use these sine basis functions because we prefer to also have a zero derivative at the beginning and end of the pulse.

\begin{figure}[t]
\includegraphics[width=0.48\textwidth]{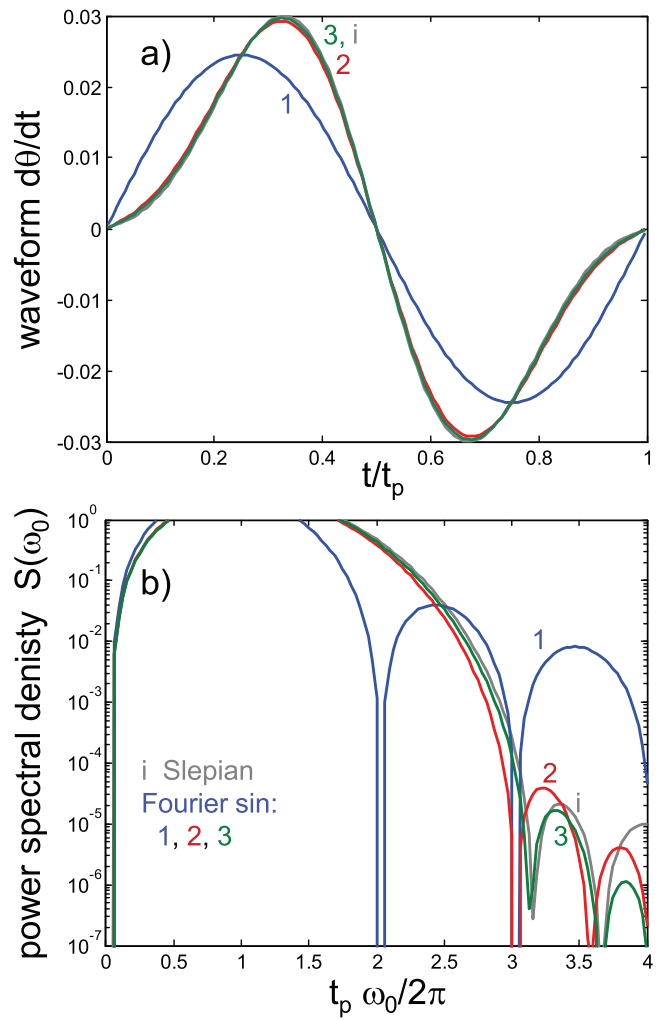}
\caption{\label{fig:DTheta} (color online) Waveform and power spectral density ($S(\omega_0) \simeq 4P_e$) as for Fig.\,\ref{fig:Theta}, but with a trajectory from $\theta_i$ to $\theta_f$ and back.  a) Plot of waveform $d\theta/dt$ versus $t/t_p$ for optimized Fourier coefficients with one (1, blue), two (2, red), and three (3, green) terms, as well as Slepian (i, gray).  b) Plot of power spectral density from Fourier transforms of the waveforms in a). Waveforms normalized using $2\sum \lambda_n' = 1$.}
\end{figure}

We are also interested in control waveforms that take an initial $\theta_i$ to $\theta_f$ and then back again to $\theta_i$.  In this case the waveform for $\theta$, not $d\theta/dt$, is given by the Fourier basis
\begin{align}
\theta - \theta_i &= \sum_{n=1,2,\ldots, n_m} \lambda_n' [1-\cos(2\pi n t/t_p)] \ ,
\label{eq:dthprime}
\end{align}
with constraints on the coefficients
\begin{align}
\theta_f-\theta_i &= 2 \sum_{n \textrm{ odd}} \lambda_n' \ .
\label{eq:dthprimecon}
\end{align}
The derivative $d\theta/dt$ is now a sum of sine functions, which have non-zero derivatives at the beginning and end of the pulse.  Figure\,\ref{fig:DTheta} shows optimized results for 1, 2, and 3 Fourier coefficients, as well as for the first-harmonic solution of the Slepian.  As for the previous case, the optimized Fourier solution is close to that of the Slepian: only two coefficients (red curve) is required for close approximation to the ideal solution.

\section{Optimum solution: arbitrary $\theta$}

The optimal solution for an arbitrary change in $\theta$ will have to account for the qubit frequency $\omega$ changing during the control pulse. The solution is found by first making a coordinate transformation to an accelerating frame in time where the oscillation frequency is constant, and then using optimal waveforms as found previously.

The basic idea is to define a new frame with time $\tau$ such that the rate of change of the phase is constant.  Choosing this frequency as the constant $\omega_x=2H_x/\hbar$, the times are related by
\begin{align}
\omega_x \,d\tau = \omega(t)\, dt \ .\label{eq:wt}
\end{align}
In this frame, the error angle from Eq.\,(\ref{GLint}) becomes
\begin{align}
\theta_{mr} &=
-\int d\tau\,(d\theta/d\tau) \, \exp[-i \omega_x \tau] \ , \label{GLtauint}
\end{align}
so that the optimal shape of $d\theta(\tau)/d\tau$ can be computed for constant frequency $\omega_x$, as done in the last section.  Integration of $d\theta/d\tau$ is next used to compute the quantity $\theta(\tau)$.  The relationship between the real time $t$ and $\tau$ is finally solved by integrating Eq.\,(\ref{eq:wt})
\begin{align}
t(\tau)&=\int^{\tau} d\tau' \, \omega_x/\omega(\tau') \\
&= \int^\tau d\tau'\, \sin\theta(\tau') \ ,
\end{align}
where in the last equation we have used Eqs.\,(\ref{eq:theta}) and (\ref{eq:omega}) to find the relation $\omega=\omega_x/\sin\theta$.
The function $t(\tau)$ can be inverted numerically using interpolation.  For example, in MATLAB code the dependence $\theta(t)$ would be given through the interpolation of the vectors $t(\tau)$ and $\theta(\tau)$ using the command  $\theta(t)$ =\textsf{interp1}$(t(\tau),\theta(\tau),t)$.

\begin{figure}[b]
\includegraphics[width=0.48\textwidth]{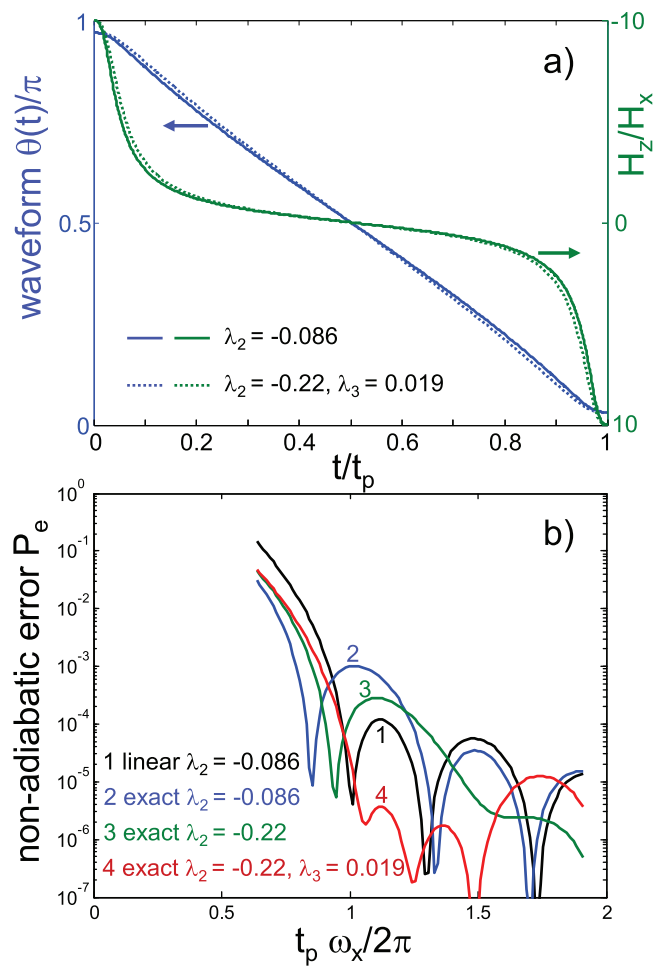}
\caption{\label{fig:LZopt} (color online) Waveform and error for $H_z$ changing from $-10H_x$ to $10H_x$.  a) Blue lines are plot of control $\theta(t)$ versus $t/t_p$ for two (solid) and three (dashed) Fourier coefficients.  The right scale shows the waveform $H_z(t)$ as green lines.  b) Plot of non-adiabatic error versus pulse time $t_p$ for linearized and exact solutions, with both two and three optimized coefficients. Low errors can be achieved for times greater than  $t_p = 2\pi/\omega_x$, corresponding to one oscillation at the avoided level crossing. We define $\omega_x=2H_x/\hbar$.}
\end{figure}

To illustrate this solution for an arbitrary control problem, we first consider the avoided level crossing from $H_z=-10H_x$ to $ H_z=10H_x$, as discussed earlier for the Landau-Zener trajectory.  The waveform is shown as solid lines in Fig.\,\ref{fig:LZopt}a for the 2 Fourier coefficients $\{ \lambda_1 = 1.086, \lambda_2=-0.086 \}$ found earlier.  Here we find the control waveform is nearly linear in $\theta$, not in $H_z$ as assumed for the standard Landau-Zener problem.  The non-adiabatic error $P_e$ is plotted in Fig.\,\ref{fig:LZopt}b, both for the linearized (black) and the exact (blue) solution of the qubit dynamics.  The linearized solution has similar shape to that found earlier, as expected, whereas the exact solution shows significantly higher errors, demonstrating that an exact solution for this control problem is important.

\begin{figure}[b]
\includegraphics[width=0.48\textwidth]{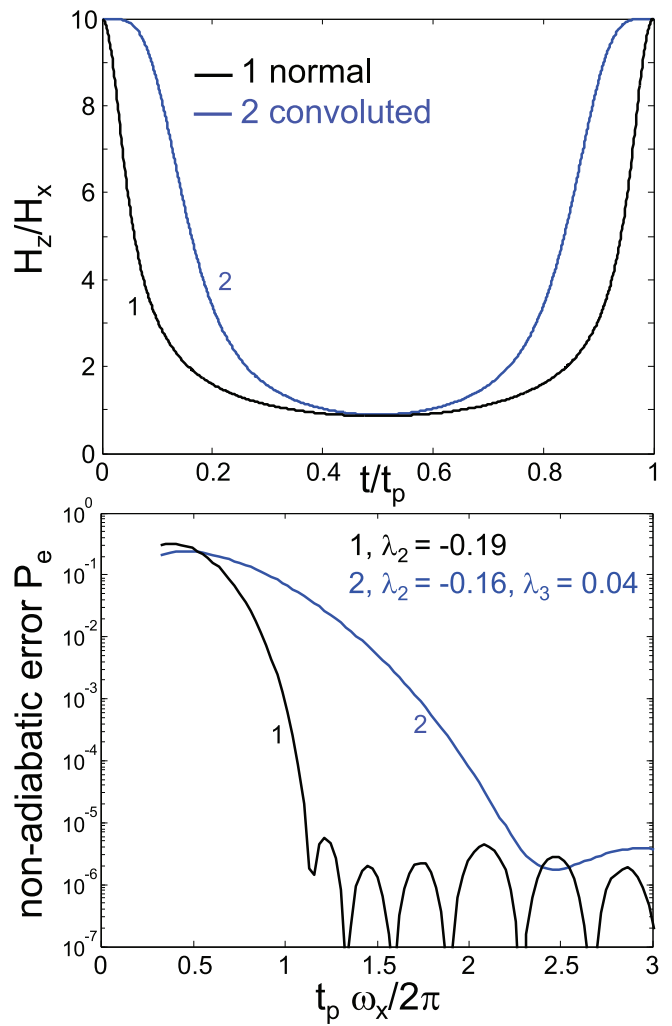}
\caption{\label{fig:CZ} (color online) Waveform and error for $H_z$ changing from $10\,H_x$ to $1.17\,H_x$, and then back to $10H_x$.  a) Black line (1) is for optimal waveform, whereas blue line (2) includes the effect of waveform rounding by control electronics, here accounted for by a convolution.  b) Plot of non-adiabatic error versus pulse time $t_p$ for the two cases, using exact solutions.  Optimal coefficients are $\{ \lambda_2', \lambda_3' \} = \{-0.19,0\}$ for black (1) and $\{-0.16,0.04\}$ for blue (2) line. }
\end{figure}

To further reduce error, we next optimize the $\lambda_n$ coefficients for better performance.  Shown in green is the best performance obtained for varying $\lambda_2$, whereas the red curve shows optimal performance for a choice of $\lambda_2$ and $\lambda_3$.  It is impressive that non-adiabatic errors in the $10^{-5}$ range can be obtained with only a small number of coefficients.  Also note that that low errors occurs at a pulse time $t_p$ equal to one oscillation period $h/2H_x$ of the avoided level crossing, about 20 times faster than the simple Landau-Zener trajectory.

Another important waveform is for a controlled pulse in $\theta$, as shown earlier in Fig.\,\ref{fig:DTheta}.  Here we consider the control angle changing from $\theta_i = 0.1$ to $\theta_f = 0.55\,\pi/2$, then back again to $\theta_i$; these parameters are chosen to give the CZ gate $|11\rangle \rightarrow \exp(i\pi)|11\rangle$, where this additional $\pi$ phase shift comes from the avoided level crossing between the states $|11\rangle$ and $|02\rangle$.  In this case $\theta(\tau)$ is given by Eqs.\,(\ref{eq:dthprime}) and (\ref{eq:dthprimecon}), and $\theta(t)$ is found as for the last example.  We find a near optimal solution for only one additional coefficient $\lambda_2' = -0.19$, as shown in black in Fig.\,\ref{fig:CZ}.  We obtain very low error for a total control time about one oscillation of $\omega_x/2\pi$.

This waveform requires fast changes in $H_x$ at the beginning and end of the pulse.  Because of the finite bandwidth of the control electronics, such a shape may not be possible to synthesize, so we next consider the optimal control including rounding of the waveform, here accounted for by convolution of the output waveform with a Gaussian.  With convolution, the computed non-adiabatic error increases significantly with the original $\lambda_2'$ coefficient.  We found that adjusting $\lambda_2'$ could not improve the error much, but by including the next coefficient $\lambda_3'$ the error could be effectively minimized.  This optimal solution is shown as the blue line in Fig.\,\ref{fig:CZ}, where low error is now found at about twice the oscillation period of $\omega_x/2\pi$.  Note that about 1/2 of the increase in time of the control waveform is accounted for by the extra time added at the beginning and end of the pulse from convolution.

\section{Optimum solution for 2-state errors}

The Fourier basis may be used to optimize waveforms for other control problems, so it is natural to ask whether these ideas can be used to minimize the excitation to the second excited state when driving weakly anharmonic qubits.  This corresponds to the standard problem for optimal control, where shaping occurs in the off-diagonal $\sigma_x$ and $\sigma_y$ parts of the Hamiltonian.  A solution that works extremely well is waveform shaping by Derivative Removal by Adiabatic Gate (DRAG) \cite{drag,gambetta}.  It is interesting to consider whether simple shaping using the Fourier basis waveforms can improve upon DRAG.

Errors from the $|2\rangle$ state come in two forms: (I) direct transitions to $|2\rangle$ producing 2-state errors, and (II) virtual excitations to $|2\rangle$, producing an AC Stark shift between the qubit states $|0\rangle$ and $|1\rangle$ that give changes in rotation amplitude and direction.  The transition frequencies between the states $|0\rangle \leftrightarrow |1\rangle$ and $|1\rangle \leftrightarrow |2\rangle$ are $\omega_0$ and $\omega_0+\Delta$, which defines a nonlinearity $\Delta$ that is negative for transmons.

(I) Transitions to the 2-state come from the drive waveform having spectral density at the $|1\rangle \leftrightarrow |2\rangle$ frequency.  As the drive needs to be resonant with the $|0\rangle \leftrightarrow |1\rangle$
transition, it can be described in general with a two-quadrature form  $x(t)\cos(\omega_0 t)+y(t)\sin(\omega_0 t)$, which can be written compactly in complex notation as $[x(t)-iy(t)]\exp(i\omega_0 t)$.  The drive waveform for DRAG has the quadrature component proportional to the time derivative $\dot{x}(t)$
\begin{align}
W(t)=[x(t)-iD\, \dot{x}(t)/\Delta]e^{i\omega_0 t} \ ,
\end{align}
with its magnitude adjusted by a dimensionless derivative parameter $D$.  The transition rate is expected to scale as the spectral density of $W$, which is the squared magnitude of the Fourier transform
\begin{align}
W(\omega) &= \int dt \, e^{-i\omega t} W(t) \\
&= [1+D(\omega-\omega_0)/\Delta] \int dt \ e^{-i\omega t} x(t)e^{i\omega_0 t} \ ,
\end{align}
where integration by parts is used \cite{frank} to shift the derivative of $x$ to the term $\exp[i(\omega_0 -\omega)t]$.  Driving the $|0\rangle \leftrightarrow |1\rangle$ transition at $\omega_0$, the spectral density is zero at the $|1\rangle \leftrightarrow |2\rangle$ transition frequency ($\omega-\omega_0=\Delta$) when $D=-1$.  As this is a qualitative argument, we use numerical simulation to minimize the population error to the 2-state; below we show our optimal DRAG coefficient is $D=-1.20$, close to the computed value.

(II) The origin of the AC Stark shift can be understood as coming from level repulsion between the $|1\rangle$ and $|2\rangle$ states when driven off-resonance, which changes the frequency of $|1\rangle$ state proportional to $-x^2/2\Delta$.  The derivative term also Stark shifts the qubit, giving a net shift \cite{ape} proportional to $-(1+2D)x^2/2\Delta$.  As the $x^2$ term gives a shift that is time dependent, a practical solution is to choose $D=-1/2$ for no Stark shift, allowing the qubit frequency to be constant \cite{ape}.  In this case transition errors to the 2-state are not optimized, but the solution is still useful since errors are lowered by $(1+D)^2 = 1/4$.

As we wish to use $D=-1$ for minimum 2-state errors, we note here there exists another practical solution: constant detuning of the microwave frequency.  This works because the dynamic Stark shift, which produces a small change in the rotation direction of the Bloch vector in the qubit subspace, can always be compensated for by a static change in drive frequency and phase.  Equivalently stated, the three parameters of the drive frequency, phase and amplitude moves the Bloch vector by any desired rotation direction and angle.  Finding zero error thus becomes the same procedure as for normal tuning up of qubit gates, only here the frequency and phase must be separately tuned for $\pi/2$ and $\pi$ rotations.  As this optimization can always be performed, we need only be concerned with 2-state errors.

\begin{figure}[t]
\includegraphics[width=0.48\textwidth]{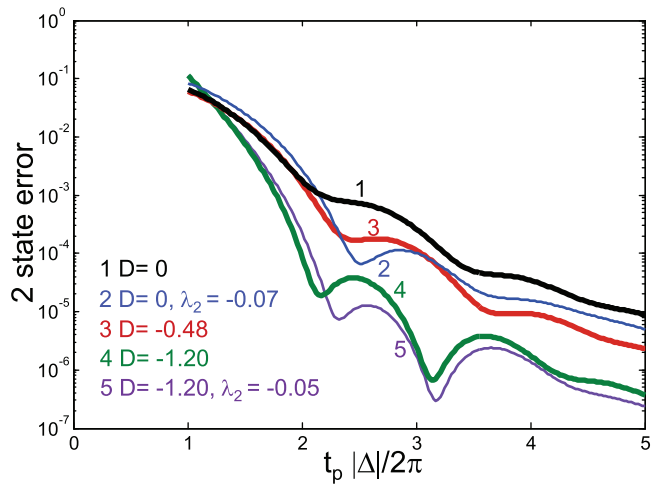}
\caption{\label{fig:HD} (color online) Average probability of two-state population versus pulse time $t_p$ for a qubit $\pi$-pulse.  The time is normalized with the qubit nonlinearity $|\Delta|/2\pi$.  Parameters are chosen to minimize qubit errors near $t_p|\Delta|/2\pi = 2.5$.  DRAG term $D=$ 0, -0.48 and -1.20 are shown in black (1), red (3) and green (4), respectively.  Optimization of the waveform $x(t)$ with Fourier coefficient $\lambda_2$ are shown in blue (2) and purple (5).}
\end{figure}

These ideas were evaluated using numerical simulations of a 3-level system with $\Delta$ nonlinearity, integrating the Hamiltonian matrix over the waveform pulse to find the net unitary transformation.  We find good performance using the Fourier basis, shown initially for a drive $x(t) \propto 1-\cos(2\pi t/t_p)$ with amplitude adjusted to give a $\pi$-pulse.  As we find transition probabilities for $|0\rangle \rightarrow |2\rangle$ and $|1\rangle \rightarrow |2\rangle$ are quite close in magnitude, we plot in Fig.\,\ref{fig:HD} the average of these 2-state errors versus pulse time $t_p$ normalized with the qubit anharmonicity $|\Delta|/2\pi$.  Here, we optimize amplitude and detuning parameters for qubit subspace errors below $10^{-7}$ at $t_p |\Delta|/2\pi \simeq 2.5$; we find the 2-state errors change little when slightly varying these parameters to zero errors at different values of $t_p$.  The error without a derivative $D=0$ is plotted in black.  The waveform with the half-derivative $D=-0.48$, plotted in red, shows a reduction in error by about a factor of 4 that is consistent with the above argument.  As expected, we find a much better solution using the DRAG coefficient near unity $D=-1.20$, as plotted in green.

We also tested for waveform optimization of $x(t)$ by adjusting the Fourier coefficient $\lambda_2$.  For the $D=0$ case we find better performance for $\lambda_2=-0.07$, plotted in blue, with an average magnitude similar to the $D=-0.48$ solution.  For $D=-1.20$ and $\lambda_2=-0.05$, plotted in purple, we found adjusting $\lambda_2$ shifted the first minimum; we thus propose that this parameter can be used to tune for a somewhat smaller 2-state error.

Comparing to plots in Ref.\,\cite{gambetta} using a truncated Gaussian waveform, we find the $D=-0.48$ solution has errors comparable to first order DRAG (purple curve of their Fig.\,3).  Our $D=-1.20$ solution is about one order of magnitude below the second order DRAG solution (green curve of their Fig.\,4) for $t_p |\Delta|/2\pi \lesssim 3.1$.  We conjecture that combining second-order DRAG with our optimized Fourier-based waveforms would perform even better.

Waveforms corrected by a derivative term with $D=-1.20$ produce a good optimum solution.  However, one-component optimal control is not that much slower, acceptable for problems where there is only access to $\sigma_z$ control.

\section{conclusion}

We have shown that an optimal control waveform for adiabatic behavior may be obtained by minimizing the spectral weight integrated above a cutoff frequency.  Although the best solution is a Slepian, optimization of a few-term Fourier series shows comparably low errors, with acceptable performance achieved with only two or three coefficients.  This solution can be used for arbitrary control, even when the transition frequency is changing in time, by first solving the problem with a fixed frequency and then transforming to the proper time frame.  We find effects of finite bandwidth in the control waveform can be readily accounted for by optimizing the Fourier coefficients.  This method of solution can be used for other control problems, such as minimizing the 2-state occupation for a single qubit $\pi$-pulse.

With this protocol, we find that gates with intrinsic error below $10^{-4}$ should be possible for the two-qubit controlled-phase operation, even for relatively fast gate times.  This theoretical understanding may help open up a route to ultra-high fidelity quantum operations and fault tolerant quantum computation.

We thank A. Korotkov for helpful discussions. This work was supported
by the Office of the Director of National Intelligence (ODNI), Intelligence Advanced Research Projects Activity (IARPA), through the Army Research Office grants W911NF-09-1-0375 and W911NF-10-1-0334. All statements of fact, opinion or conclusions contained herein are those of the authors and should not be construed as representing the official views or policies of IARPA, the ODNI, or the U.S. Government.

\section{Appendix I: Exact solution}

The exact quantum solution to the non-adiabatic problem may be expressed in the form found geometrically.  We start by considering the effect of a rotation around the $H_x$-axis.  If the wavefunction in a fixed basis is given by $|\Psi\rangle = a|0\rangle + b |1\rangle$, then it can be rewritten in a basis rotated by the angle $\theta$ as $|\Psi\rangle = \alpha |-\rangle + \beta |+\rangle$, re-expressed in eigenstates of the new basis with amplitudes given by
\begin{align}
\alpha &= a \cos(\theta/2) - b \sin(\theta/2) \ ,\\
\beta &= b \cos(\theta/2) + a \sin(\theta/2) \ .
\end{align}

In the $\theta$-rotated basis we define the energy eigenvalues as $\pm \omega/2$, which introduces phase factors that has differential form $\dot{\alpha} = -i (\omega/2)\alpha$ and $\dot{\beta} = i (\omega/2) \beta$.  We are interested in changes in the control vector, so if $\theta$ changes with time one finds
\begin{align}
\dot{\alpha} &= -i(\omega/2)\alpha-a\sin(\theta/2)\,\dot{\theta}/2 -b\cos(\theta/2)\,\dot{\theta}/2 \\
&= (-i\omega\alpha-\dot{\theta}\beta )/2 \ ,\\
\dot{\beta} &= (+i\omega\beta + \dot{\theta}\alpha )/2 \ .
\end{align}
Note the symmetry of the rotations, with factors $i\omega$ for the time-dependence of $H_z$ rotations, and $\dot{\theta}$ for the $H_x$ rotation.

The time dependence can be described simply if one considers the quantity $\alpha^*\beta$, where $^*$ is the complex conjugate.  Its derivative is
\begin{align}
\frac{d}{dt}\big(\alpha^* \beta\big) &= (i\omega\alpha^*-\beta^* \dot{\theta})\beta/2 +
\alpha^*(i\omega\beta + \alpha \dot{\theta})/2 ] \\
&= i\omega (\alpha^* \beta)+(|\alpha|^2-|\beta|^2)\dot{\theta}/2 \\
&= i\omega (\alpha^* \beta)+\sqrt{1-4|\alpha^*\beta|^2}\ \dot{\theta}/2 \ ,\label{LZde}
\end{align}
where in the last equation we have used the relation
\begin{align}
\big(|\alpha|^2-|\beta|^2\big)^2 & = |\alpha|^4+2|\alpha|^2|\beta|^2+|\beta|^4 - 4|\alpha^*\beta|^2 \\
& = \big(|\alpha|^2+|\beta|^2\big)^2 - 4|\alpha^*\beta|^2 \\
& = 1 - 4|\alpha^*\beta|^2 \ .
\end{align}
Note that the square root factor in Eq.\,(\ref{LZde}) changes sign when $|\alpha|^2-|\beta|^2$ is negative.

For Bloch vector angles $\Theta$ and $\varphi$ the amplitudes are $\alpha=\cos(\Theta/2)$ and $\beta=\sin(\Theta/2)e^{i\varphi}$, which gives
\begin{align}
\alpha^*\beta &= \cos(\Theta/2)\sin(\Theta/2)e^{i\varphi} \\
&=\sin\Theta \,e^{i\varphi} /2 \ .
\end{align}
so that its magnitude is the error probability $P_e = [\sin(\Theta/2)]^2 \approx |\alpha^*\beta|$ only for small angles $\Theta$. Inserting this result for $\alpha^*\beta$ into Eq.\,(\ref{LZde}), one finds
\begin{align}
\frac{d}{dt}(\sin\Theta \,e^{i\varphi}) = i\omega (\sin\Theta \,e^{i\varphi}) + \cos \Theta\  \dot{\theta} \ .
\end{align}
Changing phase variables $\varphi = \varphi' + \phi$, where $\phi = \int \omega dt$ and noting that $d\phi/dt = \omega$, we can remove the $i\omega$ phase term
\begin{align}
\frac{d}{dt}(\sin\Theta \,e^{i\varphi'}) = \cos \Theta\  \dot{\theta} e^{-i\phi} \ ,
\end{align}
which is equivalent to previous geometrical result Eq.\,(\ref{GLint}) in the low amplitude limit $\Theta \rightarrow 0$.  Here, the exact solution has the amplitude of the drive term $d\theta /dt$ rescaled by $\cos\Theta$ to account for the geometry of the Bloch sphere.


\begin{thebibliography}{10}
\bibitem{gdef} The interacting Hamiltonian is $H_\textrm{int} = \hbar g(|10\rangle\langle 01|+|01\rangle\langle 10|)$.  The coupling between states $|11\rangle$ and $|02\rangle$ is larger by approximately $\sqrt{2}$.
\end{thebibliography}
\end{document}